# Complex magnetic order in the kagomé staircase compound $Co_3V_2O_8$

Y. Chen
*NIST Center for Neutron Research, National Institute of Standards and Technology, Gaithersburg, Maryland 20899, USA and Department of Materials Science and Engineering, University of Maryland, College Park, Maryland 20742, USA*

J. W. Lynn, Q. Huang, F. M. Woodward, and T. Yildirim
*NIST Center for Neutron Research, National Institute of Standards and Technology, Gaithersburg, Maryland 20899, USA*

G. Lawes
*Department of Physics and Astronomy, Wayne State University, Detroit, Michigan 48201, USA*

A. P. Ramirez
*Bell Labs, Lucent Technologies, 600 Mountain Avenue, Murray Hill, New Jersey 07974, USA*

N. Rogado
*Department of Chemistry and Princeton Materials Institute, Princeton University, Princeton, New Jersey 08544, USA and DuPont Central Research and Development, Experimental Station, Wilmington, Delaware 19880-0328, USA*

R. J. Cava
*Department of Chemistry and Princeton Materials Institute, Princeton University, Princeton, New Jersey 08544, USA*

A. Aharony and O. Entin-Wohlman
*School of Physics and Astronomy, Raymond and Beverly Sackler Faculty of Exact Sciences, Tel Aviv University, Tel Aviv 69978, Israel and Department of Physics, Ben Gurion University, Beer Sheva 84105, Israel*

A. B. Harris
*Department of Physics and Astronomy, University of Pennsylvania, Philadelphia, Pennsylvania 19104, USA*

$Co_3V_2O_8$ (CVO) has a different type of geometrically frustrated magnetic lattice, a kagomé staircase, where the full frustration of a conventional kagomé lattice is partially relieved. The crystal structure consists of two inequivalent (magnetic) Co sites, one-dimensional chains of Co(2) spine sites, linked by Co(1) cross-tie sites. Neutron powder diffraction has been used to solve the basic magnetic and crystal structures of this system, while polarized and unpolarized single crystal diffraction measurements have been used to reveal a rich variety of incommensurate phases, interspersed with lock-in transitions to commensurate phases. CVO initially orders magnetically at 11.3 K into an incommensurate, transversely polarized, spin density wave state, with wave vector $k=(0,\delta,0)$ with $\delta=0.55$ and the spin direction along the $a$ axis. $\delta$ is found to decrease monotonically with decreasing temperature and then locks into a commensurate antiferromagnetic structure with $\delta=\frac{1}{2}$ for $6.9<T<8.6$ K. In this phase, there is a ferromagnetic layer where the spine site and cross-tie sites have ordered moments of 1.39 $\mu_B$ and 1.17 $\mu_B$, respectively, and an antiferromagnetic layer where the spine-site has an ordered moment of 2.55 $\mu_B$, while the cross-tie sites are fully frustrated and have no observable ordered moment. Below 6.9 K, the magnetic structure becomes incommensurate again, and the presence of higher-order satellite peaks indicates that the magnetic structure deviates from a simple sinusoid. $\delta$ continues to decrease with decreasing temperature and locks in again at $\delta=\frac{1}{3}$ over a narrow temperature range ($6.2<T<6.5$ K). The system then undergoes a strongly first-order transition to the ferromagnetic ground state ($\delta=0$) at $T_c=6.2$ K. The ferromagnetism partially relieves the cross-tie site frustration, with ordered moments on the spine-site and cross-tie sites of 2.73 $\mu_B$ and 1.54 $\mu_B$, respectively. The spin direction for all spins is along the $a$ axis (Ising-like behavior). A dielectric anomaly is observed around the ferromagnetic transition temperature of 6.2 K, demonstrating that there is significant spin-charge coupling present in CVO. A theory based on group theory analysis and a minimal Ising model with competing exchange interactions can explain the basic features of the magnetic ordering.

## I. INTRODUCTION

The ground state properties of geometrically frustrated magnetic systems, where frustration arises from incompatibility of the local antiferromagnetic (AF) interactions with the global symmetry imposed by the crystal lattice, have been a subject of recent intensive experimental and theoretical work. A large number of different ground states have been discovered, such as “spin glass,” “spin ice,” and numer-ous spin liquidlike states.[1,2] Recently, geometrically frustrated magnetic systems have gained greater interest as new exciting materials, multiferroics, which exhibit a coupling of the magnetic order with the dielectric properties while having a magnetic lattice that is frustrated.[3–13] For example, the Mn ions in $HoMnO_3$ form a triangular lattice whose magnetic phase can be controlled by an external electric field.[4,7–9,12,13]

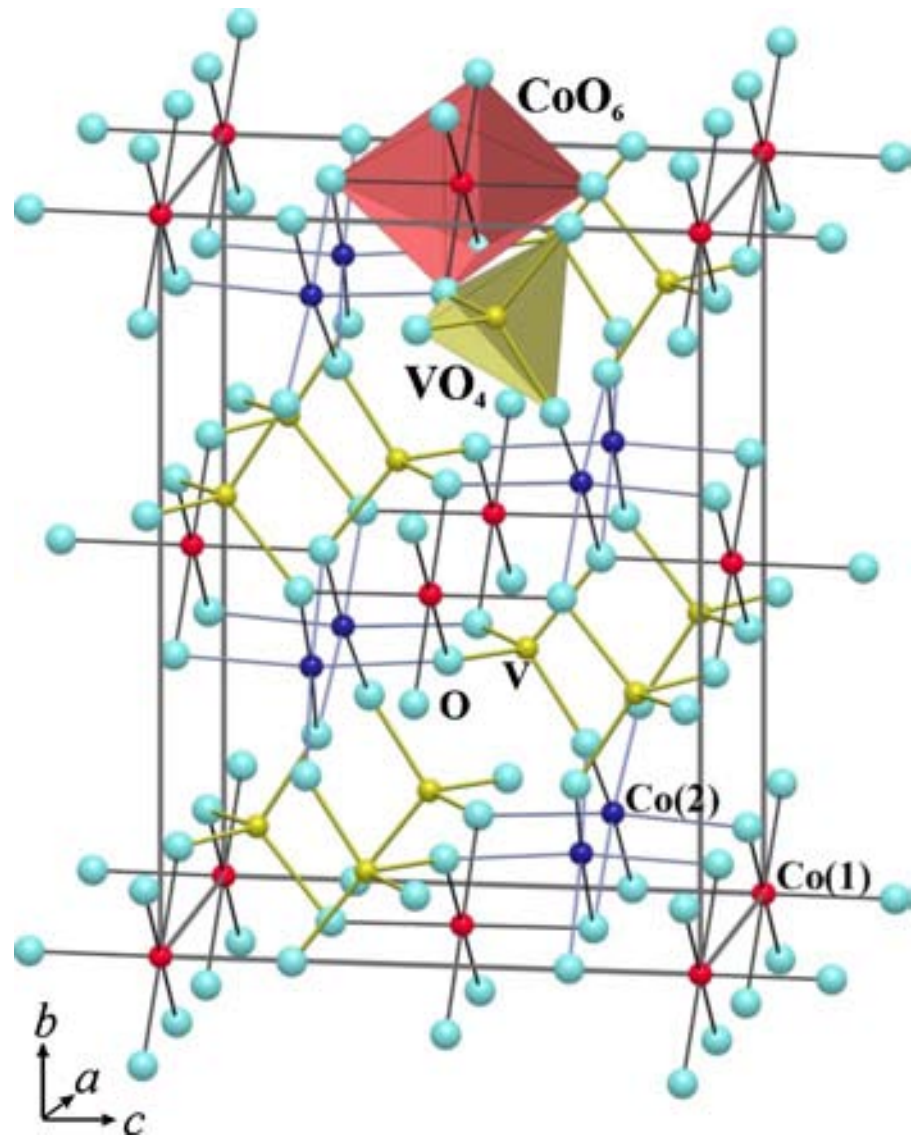

FIG. 1. (Color online) Crystal structure of $Co_3V_2O_8$. The nonmagnetic V-O tetrahedra provide isolation between the magnetic Co-O layers.

A two-dimensional kagomé lattice, consisting of corner-sharing triangles of spins with antiferromagnetic coupling between nearest neighbors, is one of the best-known frustrated lattices.[14–17] A variation of the kagomé net, the kagomé staircase lattice, has been recently realized in the $M_3V_2O_8$ ($M$=$Ni^{2+}$, $Co^{2+}$, $Zn^{2+}$, and $Cu^{2+}$) compounds.[18–23] These materials consist of buckled kagomé layers of edge-sharing transition metal oxide octahedra separated by nonmagnetic $V_5O_4$ tetrahedra. Figure 1 shows the crystal structure of $Co_3V_2O_8$ (CVO). The edge-sharing $CoO_6$ octahedral layers lie in the $a$-$c$ plane, separated by V-O tetrahedra in the $b$ direction. The nonmagnetic V-O tetrahedra provide isolation between the magnetic Co-O layers. At 15 K, the average Co-Co distance within the layers $d_1 \approx 2.99$ Å, while the interlayer distance $d_2 = b/2 \approx 5.74$ Å. The relatively large interlayer to intralayer ratio, $d_2/d_1 \approx 1.9$, and the indirect interlayer Co-O-O-Co superexchange pathway suggest a strong two-dimensional magnetic character in the compound, with the magnetism dominated by intralayer $Co^{2+}$-O-$Co^{2+}$ coupling interactions. However, unlike previously studied kagomé lattice-based materials, these corner-sharing kagomé layers in CVO [Fig. 2(a)] are not flat but instead are buckled, resulting in a kagomé staircase geometry [Fig. 2(c)]. In addition, unlike the fully frustrated conventional kagomé net, the $S = \frac{3}{2}$ $Co^{2+}$ ions have two distinct crystallographic sites: the Co(1) cross-tie and the Co(2) spine sites. The Co(2) sites form chains along the $a$ axis, which are linked by the Co(1) sites in the $c$ direction to form the buckled kagomé staircase. These planes are stacked along the $b$ direction to complete the structure [Fig. 2(c)]. The buckled kagomé layers, the resulting inequivalent superexchange interactions, and the an-isotropic magnetic coupling in the " staircase" magnetic layers contribute to the reduction of the effects of geometric frustration in CVO and result in the appearance of multiple temperature-dependent magnetic phase transitions. The isostructural $S=1$ $Ni_3V_2O_8$ (NVO) compound simultaneously develops ferroelectricity and incommensurate magnetic order, demonstrating that the two order parameters are strongly coupled.[24] A rich anisotropic magnetic field-temperature phase diagram was observed,[21] and NVO shows an amplitude-modulated high-temperature incommensurate phase and a helical low-temperature incommensurate phase.

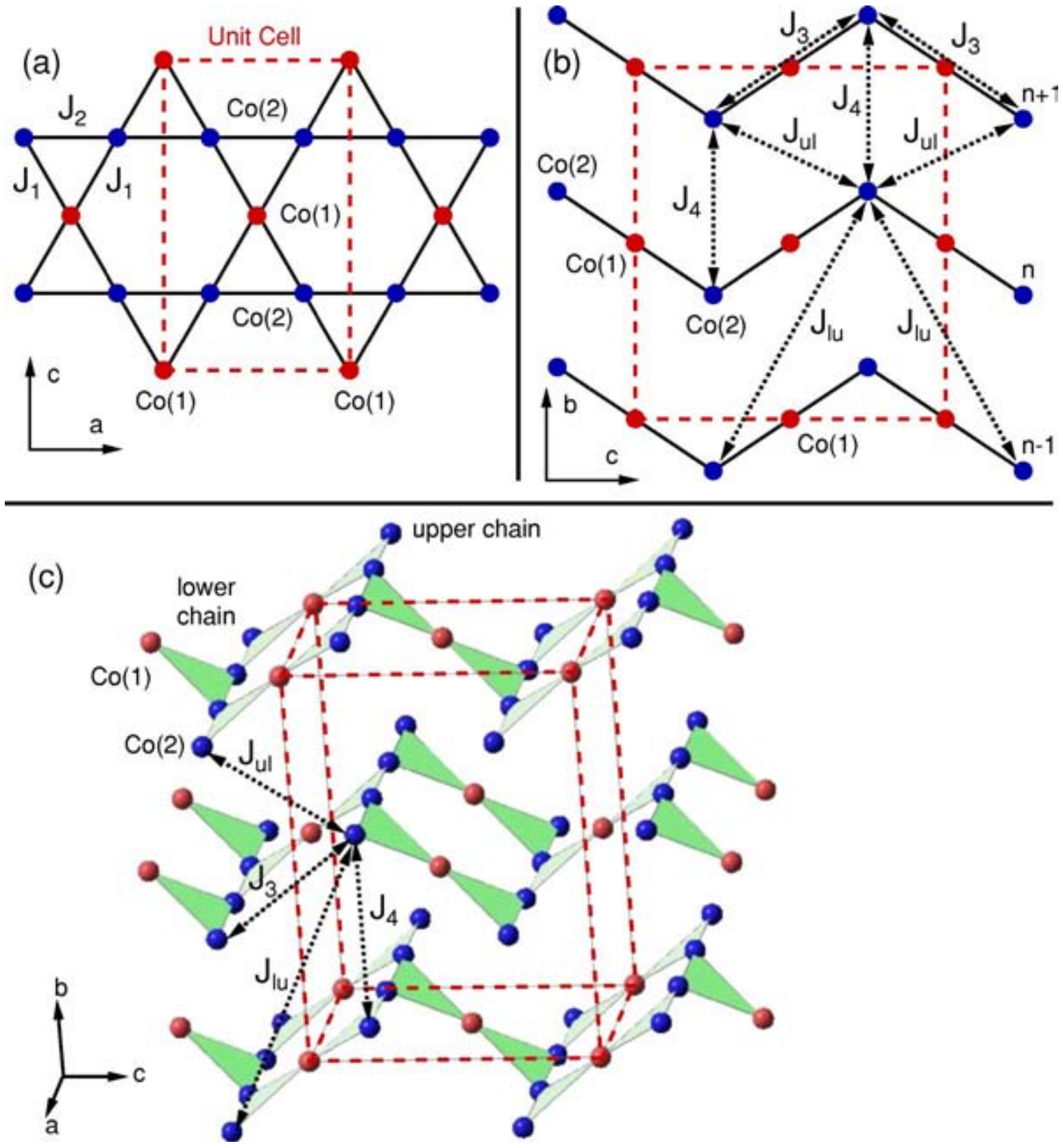

FIG. 2. (Color online) (a) The crystal structure of CVO viewed down the crystallographic $c$ axis. There are two Co sites: the Co(1) cross-tie (at the apex of the triangles) and Co(2) spine sites. (b) Edge-on view of the kagomé staircase. (c) Kagomé staircase lattice, showing only the Co sites. Note that when the Co(2) sites are ordered antiferromagnetically, the Co(1) spins are fully frustrated.

In this paper, powder and single crystal neutron diffraction data of CVO are presented. The data show that CVO undergoes an incommensurate magnetic transition at 11.3 K with propagation vector $k=(0,\delta,0)$, where $\delta=0.55$. $\delta$ decreases as temperature decreases. The incommensurate intensity locks into a commensurate, antiferromagnetic position with $\delta=\frac{1}{2}$ for temperatures between 6.9 and 8.6 K. Below 6.9 K, $\delta$ continues to decrease with decreasing temperature and the magnetic structure becomes incommensurate again. $\delta$ locks in again at $\delta=\frac{1}{3}$ in a narrow temperature region between 6.2 and 6.5 K. Finally, the system enters a ferromagnetic ground state ($\delta=0$) at $T_c=6.2$ K. Such complex magnetic order is a direct result of the competing magnetic interactions in the system, and the phase diagram of CVO is very different from that of the isostructural NVO. The small quantitative changes in the parameters that control the release of the frustration of the system may result in the difference. A dielectric anomaly is clearly observed at the ferromagnetic transition temperature of 6.2 K, exhibiting a significant spin-charge coupling present in CVO.

## II. EXPERIMENT

### A. Sample preparation and characterization

Powder samples of CVO were synthesized with a reported technique.[18] $Co_3O_4$ and $V_2O_5$ were used as starting materials. The reagents were mixed thoroughly and heated in dense $Al_2O_3$ crucibles. The mixture was heated under $N_2$ flow at 800 °C for 12 h, and then at 900 °C for 12 h, with intermediate grindings. The resulting powder sample was pressed into a pellet and sintered under $N_2$ flow at 1000 °C for 12 h, then at 1100 °C for 12 h.

Single crystals of CVO were grown from a $K_2O$-$V_2O_5$ flux. The optimum condition for growing crystals was achieved by mixing together $K_2CO_3$:$Co_3O_4$:$V_2O_5$ in a 1.5:1:3 ratio and placing the mixture in a dense alumina crucible. The mixture was heated in a vertical tube furnace in air for an hour at 1100 °C. The melt was cooled slowly to 900 °C at 0.1 °C/min, and then left to cool in the furnace to room temperature. The dark-colored platelike crystals were then separated from the flux.

The samples were found to be single phase by powder x-ray diffraction. The crystal structure of CVO was orthorhombic (space group *Cmca*) with $a=6.045$ Å, $b=11.517$ Å, and $c=8.316$ Å at 10 K. These values are in close agreement with previously reported data.[25]

The specific heat of a 15 mg single crystal was measured on a Quantum Design PPMS using a relaxation method. The specific heat of CVO (plotted in Fig. 3 as $C/T$ vs $T$) shows three large peaks at $T=6.2$, 6.5, and 11.3 K, corresponding to three distinct magnetic phase transitions. Although initial specific heat measurements did not have the temperature resolution to identify the two-peak structure at 6.2 and 6.5 K,[18] more precise recent results find such a splitting.[22] Additionally, there is a small but distinct feature at $T \approx 8.9$ K. This small peak is reproducible and shifts in an applied magnetic field. It corresponds to a fourth magnetic phase transition in CVO, consistent with recent reported data in Ref. 22. The specific heat data on the powder sample only shows the 11.3 and 6.2 K magnetic transitions.[18] The characteristics of these four magnetic transitions are discussed below in more detail. A large magnetic anisotropy in the compound is reported.[20,22] The direction of the easy-axis magnetization is along the $a$ axis, while the hard axis is the $b$ axis. Magnetic field-induced transitions are observed for various orientations of the magnetic field.[22]

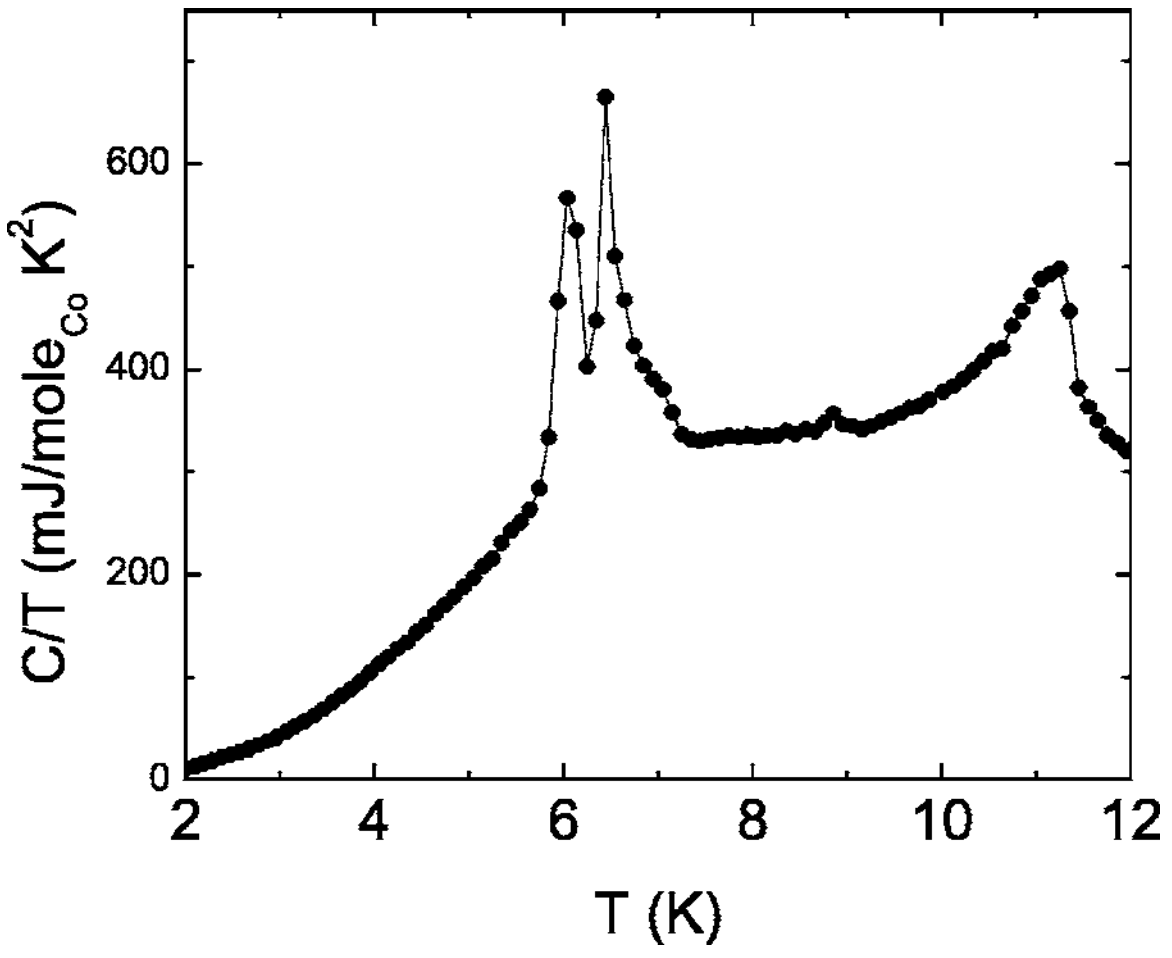

FIG. 3. The specific heat of a single crystal sample of CVO as a function of temperature is plotted as $C/T$ vs $T$, at zero magnetic field.

### B. Neutron diffraction measurements

The neutron powder diffraction data of CVO were collected on the BT-1 high-resolution powder diffractometer at the NIST Center for Neutron Research, using Cu(311) and Ge(311) monochromators to produce monochromatic neutron beams of wavelength 1.5401 and 2.0775 Å, respectively. Collimators with horizontal divergences of 15′-20′-7′ were used before and after the monochromator, and after the sample, respectively. Measurements were taken for scattering angle $2\theta$ between 3° to 168° in steps of 0.05°. Data were collected at temperatures from 300 to 3 K using a closed-cycle refrigerator to search for magnetic and possible crystal structure transitions.

Single crystal and additional powder diffraction data were obtained at the former BT-2 and BT-7 triple-axis spectrometers at the NIST Center for Neutron Research. At BT-7, a double-crystal pyrolytic graphite PG(002) monochromator was employed at a fixed incident energy of $E=13.4075$ meV ($\lambda=2.47$ Å), with 40′ collimation after the sample and no analyzer. At BT-2, the (0,0,2) reflection of pyrolytic graphite was used for selecting incident neutrons with $E=14.7$ meV ($\lambda=2.359$ Å). Horizontal Soller slits of 40′ were placed before and after the sample. A pyrolytic graphite (PG) filter was placed before the monochromator to remove higher-order wavelength contaminations at BT-2 and BT-7. A pumped helium cryostat was employed to achieve temperatures from above the magnetic ordering temperature to 1.5 K. Polarized neutron data were collected using BT-2 with a Heusler polarizing monochromator and analyzer at 14.7 meV.[26] Spin flippers and guide fields were used before and after the sample. Throughout the paper, we index the wave vectors in the orthorhombic reciprocal space with $a^*=2\pi/a$, $b^*=2\pi/b$, and $c^*=2\pi/c$.

### C. Dielectric measurements

The dielectric constant was measured on a 5 mm $\times$3 mm$\times$3 mm pressed powder and single crystal of CVO with silver paint electrodes. The capacitance was measured using an Agilent 4284A LCR meter at 100 kHz with a 1 V excitation, and sample temperature was controlled using a Quantum Design PPMS.

## III. RESULTS

### A. Powder diffraction

The structure refinements of CVO were carried out using the program GSAS,[27] with previously reported structural parameters[18,28] as initial values on powder diffraction data. The neutron scattering amplitudes used in the refinements were 2.53, −0.38, and 5.81 fm for Co, V, and O, respectively. The atomic position and temperature factor for V were fixed in all refinements due to the small neutron scattering amplitude of V. The crystal structure has *Cmca* symmetry. No detectable structural transitions were found between 300 and 1.5 K. The structural parameters and selected interatomic distances at several low temperatures are shown in Table I.

Figure 4 shows the observed, calculated and difference low angle portions of the neutron powder diffraction patterns of CVO at 15, 8.4, and 3.1 K. The upper curve of Fig. 4(a) shows the measured intensities (plus signs) and the intensities calculated based on the nuclear structure model (solid line) at 15 K. The short vertical lines mark the calculated Bragg peak positions. The calculated intensity agrees very well with the measured intensity, as shown by the difference pattern [lower curve of Fig. 4(a)]. At 8.4 K [Fig. 4(b)], in addition to nuclear structural peaks, magnetic Bragg peaks are observed. The magnetic peaks can be indexed in terms of an incommensurate propagation wave vector (0, $\delta$, 0). The intensity and position of the (0, $\delta$, 0) peak are plotted as a function of temperature (Fig. 5). On decreasing the temperature, an incommensurate peak first becomes apparent at 11.3 K. The peak intensity increases smoothly with decreasing temperature until 7.8 K, increases more rapidly and reaches a maximum at 6.2 K, near the onset of ferromagnetism, then decreases quickly and disappears completely below 5.5 K when the system is fully ferromagnetic. The wave vector $\delta$ depends strongly on temperature and decreases from about $\delta=0.55$ when the long-range order first appears at about 11.3 K to $\delta=0$, when the ferromagnetic ground state is fully developed below 5.5 K. We observed that $\delta$ locks into the commensurate value of $\delta=\frac{1}{2}$ and $\frac{1}{3}$. We will see below more detailed single crystal results, which show that the lock in at $\delta=\frac{1}{2}$ extends over a temperature range between 6.9 and 8.6 K, whereas the lock-in phase at $\frac{1}{3}$ occurs over a narrow range of temperature between 6.2 and 6.5 K. In the powder, this phase overlaps with the ferromagnetic phase due to the first-order nature of the transition. It also occurs in a very narrow temperature range. In addition, polarized beam results on the single crystal indicate a significant structural distortion in the $\delta=\frac{1}{3}$ phase. Hence, detailed refinements of the magnetic structure of the $\delta=\frac{1}{3}$ phase were not possible. It is worth pointing out that there are other weak higher-order satellite peaks observed between 7 and 6.2 K, indicating that the magnetic structure is distorted from a simple spin density wave.

TABLE I. Structural parameters and selected interatomic distances for CVO. Space group *Cmca* (No. 64). Atomic positions: Co(1): 4*a* (0,0,0) and Co(2): 8*e*(1/4, *y*, 1/4), V: 8*f*(0, *y*, *z*), O(1): 8*f*(0, *y*, *z*), O(2): 8*f* (0, *y*, *z*), O(3): 16*g* (*x*, *y*, *z*). *B* is the isotropic temperature factor, and $B=8\pi^2\langle u^2\rangle$, where $\langle u^2\rangle$ is mean squared displacement. $R_p$ is residual, and $R_{wp}$ is weighted residual.

| | $T$ (K) | 12.50 | 8.38 | 2.62 |
|---|---|---|---|---|
| | $a$(Å) | 6.02732(5) | 6.02683(6) | 6.0262(6) |
| | $b$(Å) | 11.4832(1) | 11.4825(6) | 11.4812(6) |
| | $c$(Å) | 8.29555(6) | 8.29594(8) | 8.29760(9) |
| Co(1) | $B$(Å$^2$) | 0.2(1) | 0.3(1) | 0.0(1) |
| Co(2) | $y$ | 0.1299(6) | 0.1299(6) | 0.1304(6) |
| | $B$(Å$^2$) | 0.23(6) | 0.16(6) | 0.16(6) |
| V | $x$ | 0.3762 | 0.3762 | 0.3762 |
| | $y$ | 0.1196 | 0.1196 | 0.1196 |
| | $B$(Å$^2$) | 0.24 | 0.24 | 0.24 |
| O(1) | $y$ | 0.2501(4) | 0.2503(3) | 0.2490(4) |
| | $z$ | 0.2306(4) | 0.2303(3) | 0.2301(2) |
| | $B$(Å$^2$) | 0.29(4) | 0.32(4) | 0.23(4) |
| O(2) | $y$ | 0.0015(4) | 0.0023(3) | 0.0008(4) |
| | $z$ | 0.2431(3) | 0.2430(3) | 0.2441(3) |
| | $B$(Å$^2$) | 0.36(3) | 0.33(3) | 0.25(4) |
| O(3) | $x$ | 0.2702(2) | 0.2703(2) | 0.2703(2) |
| | $y$ | 0.1185(1) | 0.1182(1) | 0.1184(1) |
| | $z$ | 0.9985(2) | 0.9986(2) | 0.9988(2) |
| | $B$(Å$^2$) | 0.36(3) | 0.33(3) | 0.25(4) |
| | $R_p$(%) | 3.94 | 4.05 | 4.55 |
| | $R_{wp}$(%) | 5.03 | 5.16 | 5.62 |
| | $\chi^2$ | 1.157 | 1.202 | 1.252 |
| | Selected interatomic distances (Å) | | | |
| Co(1)-O(2) | *x*2 | 2.017(2) | 2.016(2) | 2.025(3) |
| Co(1)-O(3) | *x*4 | 2.122(1) | 2.120(1) | 2.120(1) |
| Co(2)-O(1) | *x*2 | 2.050(6) | 2.051(5) | 2.067(5) |
| Co(2)-O(2) | *x*2 | 2.108(1) | 2.103(6) | 2.086(6) |
| Co(2)-O(3) | *x*2 | 2.094(2) | 2.094(2) | 2.089(2) |

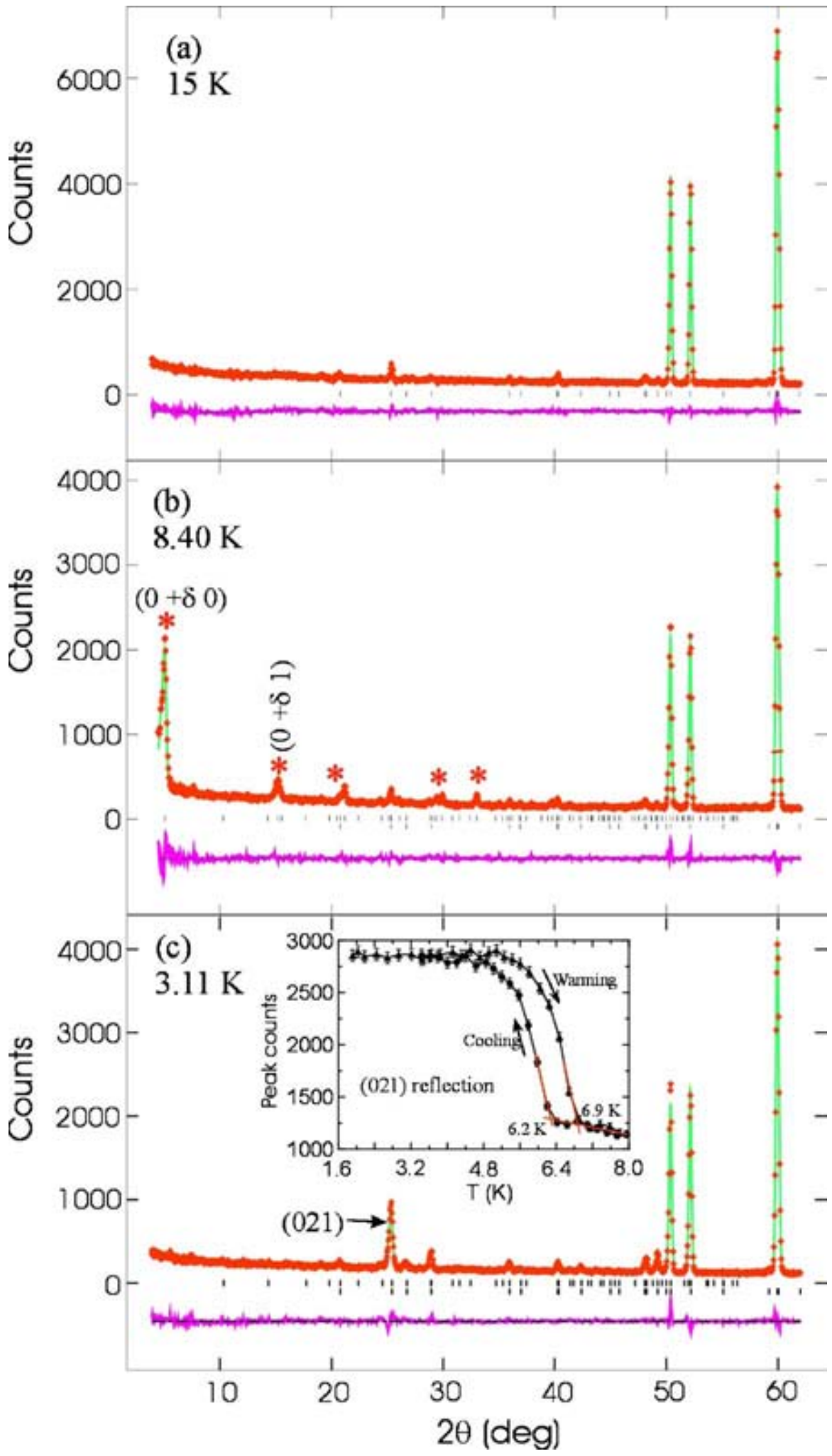

FIG. 4. (Color online) Observed (plus signs), calculated (solid lines), and difference (bottom curve) low angle portions of the neutron powder diffraction intensity. (a) $T=15$ K. Fit to a nuclear structure. (b) $T=8.4$ K. Fit includes an incommensurate magnetic structure. (c) $T=3.1$ K. Fit includes a ferromagnetic structure. The vertical lines indicate the calculated Bragg peak positions. The lower row of vertical lines in (b) and (c) indicates the nuclear Bragg peak positions.

The incommensurate magnetic structure model that best explains the observed intensities in Fig. 4(b) is a model of Co spins obeying a transversely polarized spin density wave, with wave vector $k=(0,\delta,0)$ along the $b$ axis and the spin direction along the $a$ axis, as shown in Fig. 6. Figure 7 shows the spin model for the $\delta=0.5$, commensurate, antiferromagnetically ordered phase. In this phase, the adjacent kagomé layers alternate between ferromagnetic and antiferromagnetic coupling of the ferromagnetic Co(2) chains. The moments of all the Co atoms are aligned in the crystallographic $a$-axis direction. The refined magnetic moments of Co in the ferromagnetic layer are 1.17(6) $\mu_B$ and 1.39(3) $\mu_B$ for Co(1) and Co(2), respectively, and they are 0 and 2.55(3) $\mu_B$ for Co(1) and Co(2), respectively, in the antiferromagnetic layer at 8.4 K. The Co(1) site fails to exhibit anordered moment in the antiferromagnetic layer due to frustration as a result of trying to satisfy interactions with the two antiparallel Co(2) chains, which the Co(1) sites link.

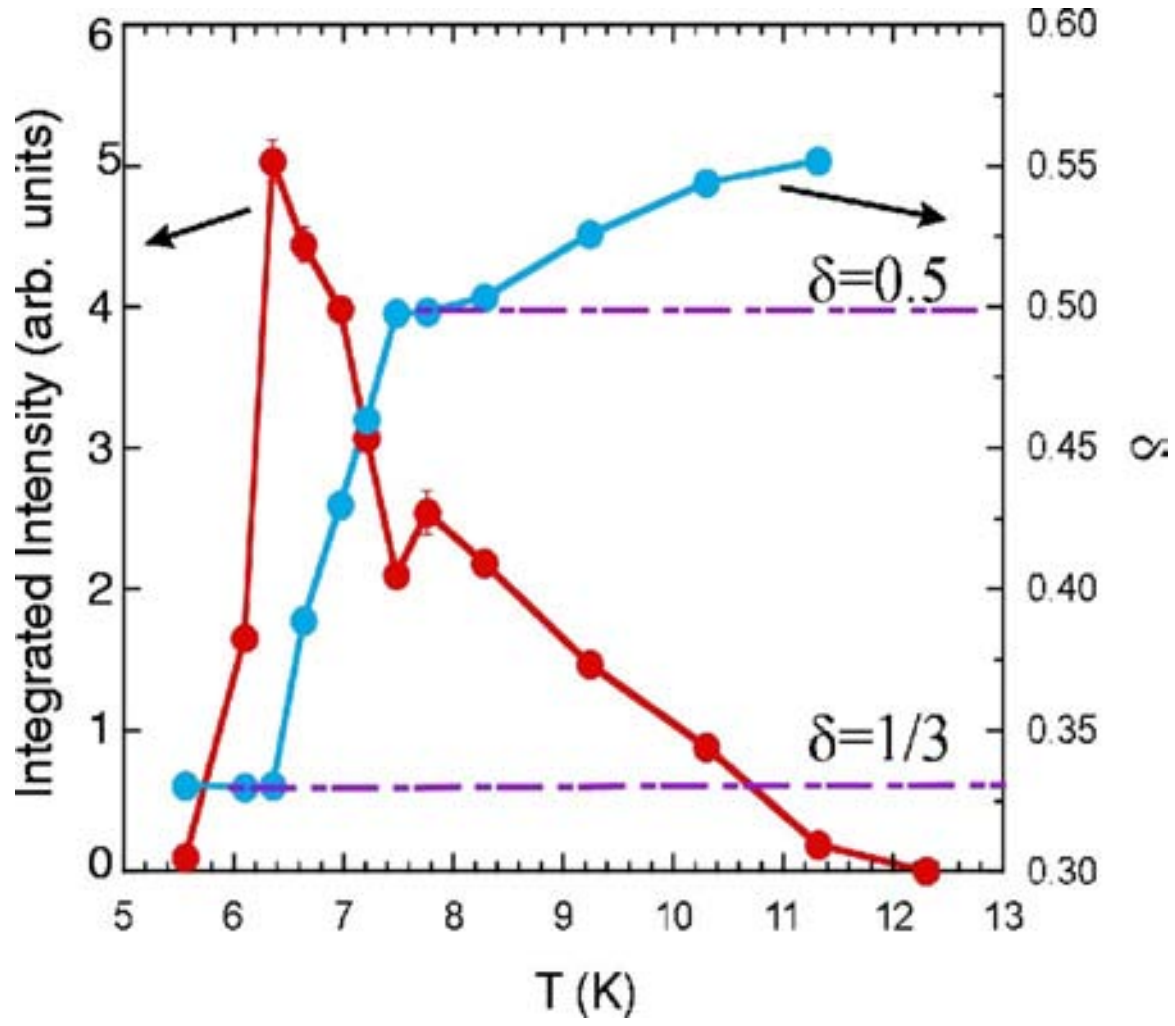

FIG. 5. (Color online) The integrated intensity (left axis) and the position (right axis) of the $(0,\delta,0)$ peak as a function of temperature measured on the powder sample of CVO.

At 3.1 K [Fig. 4(c)], a ferromagnetic model best explains the magnetic structure. Figure 8 shows the ferromagnetic structure. All the Co moments are aligned along the crystallographic *a* axis. The moments at 3.1 K are 1.54(6) and 2.73(2) $\mu_B$ for Co(1) and Co(2), respectively. The inset of Fig. 4(c) shows the temperature dependence of the (0,2,1) Bragg peak on cooling and warming. The observed hysteresis indicates that the ferromagnetic transition is first order. The ferromagnetic ground state of CVO determined directly by neutron scattering is consistent with magnetization data with the field applied along the *a* axis.[22]

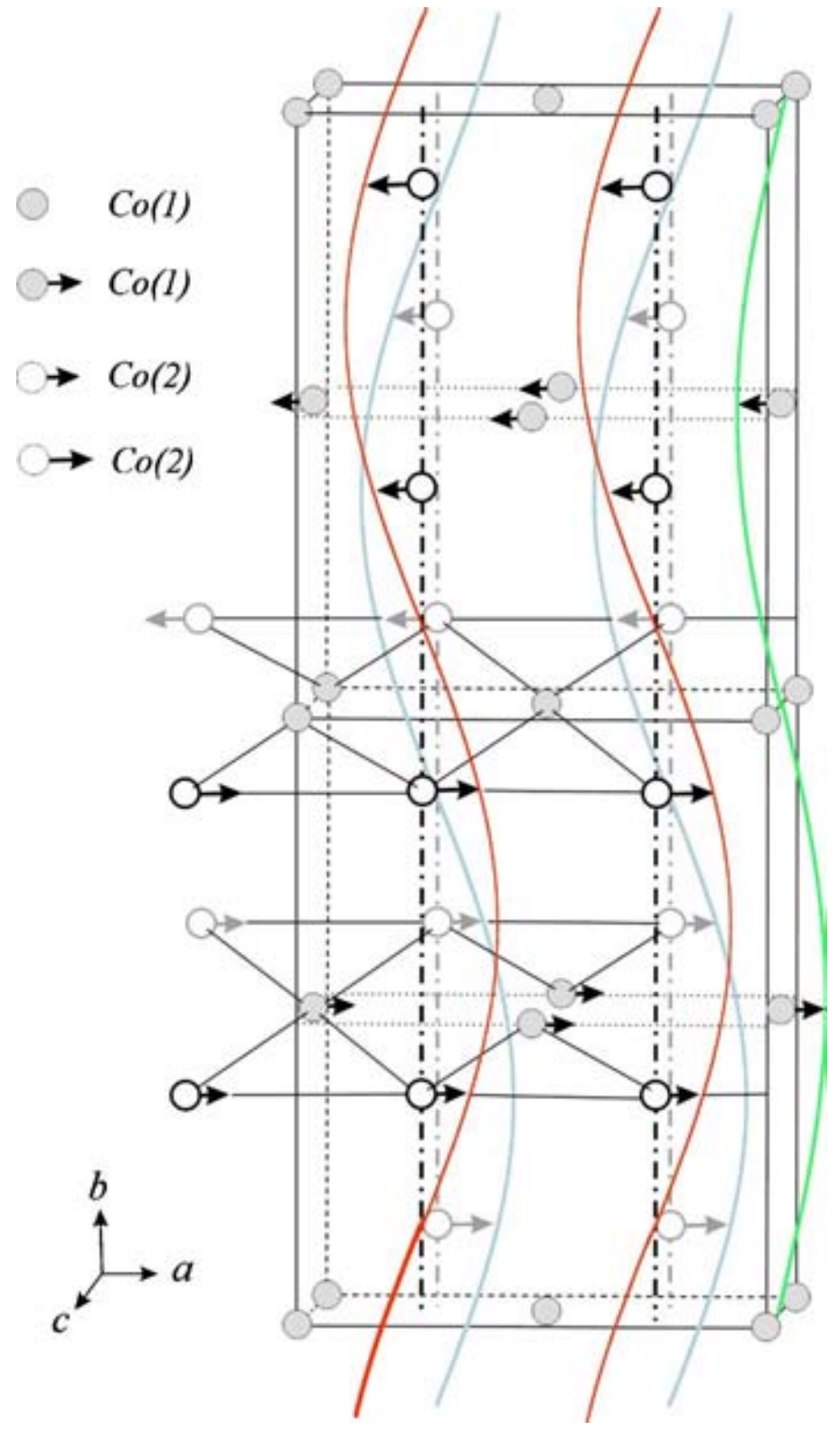

FIG. 6. (Color online) Incommensurate magnetic structure model. The transversely polarized spin density wave shown in the figure corresponds to the initial ordering wave vector of $\delta\approx 0.55$ for CVO. The wave vector is along the $b$ axis, and the direction of the moments is along the $a$ axis.

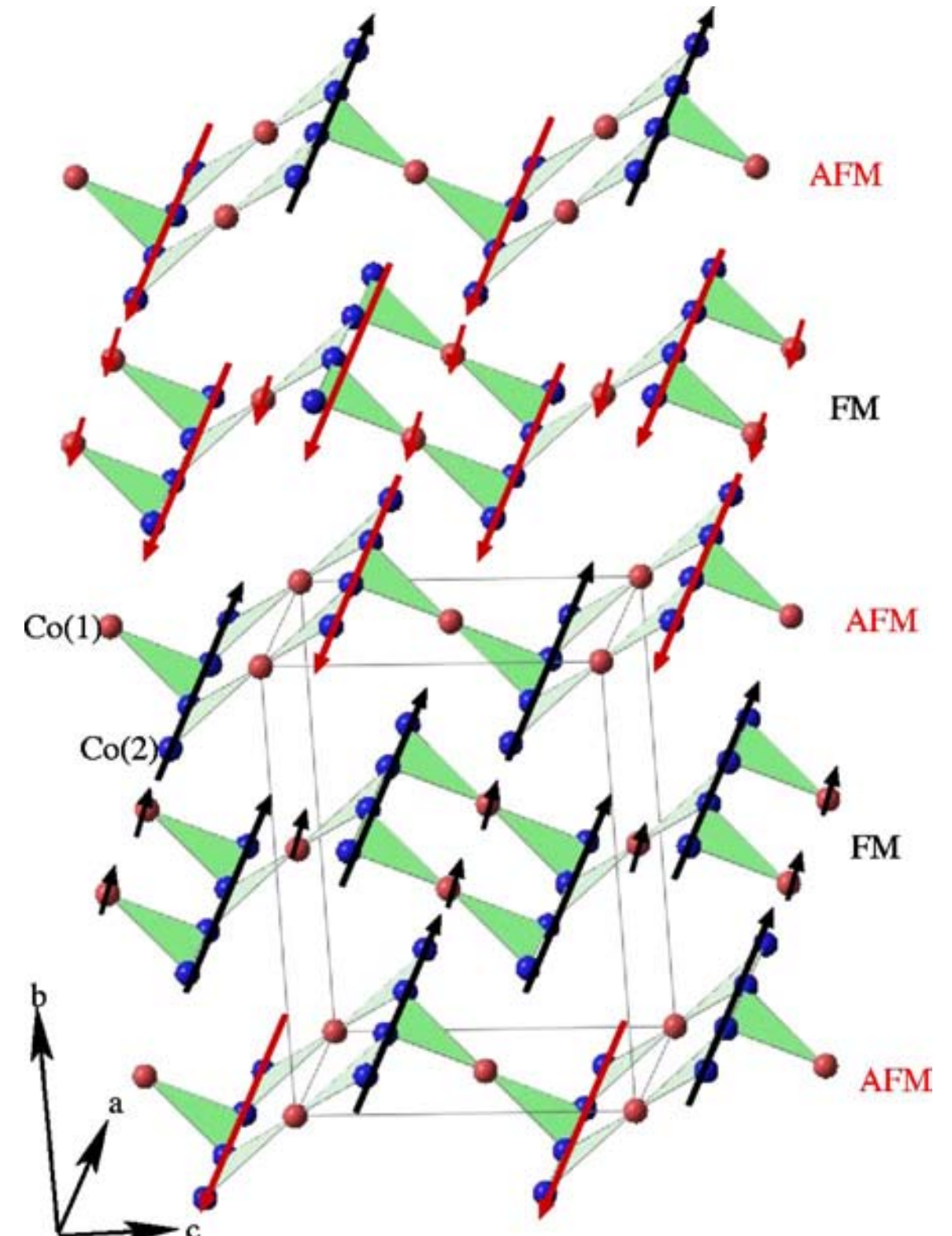

FIG. 7. (Color online) Antiferromagnetic structure model of CVO with $\delta$=0.5. The adjacent kagomé layers alternate between ferromagnetic and antiferromagnetic coupling of the ferromagnetic Co(2) chains. The Co(1) sites fail to exhibit an ordered moment in the antiferromagnetic layer due to frustration as a result of trying to satisfy interactions with the two antiparallel Co(2) chains, which Co(1) sites link. The direction of the Co moments is along the crystallographic *a* axis.

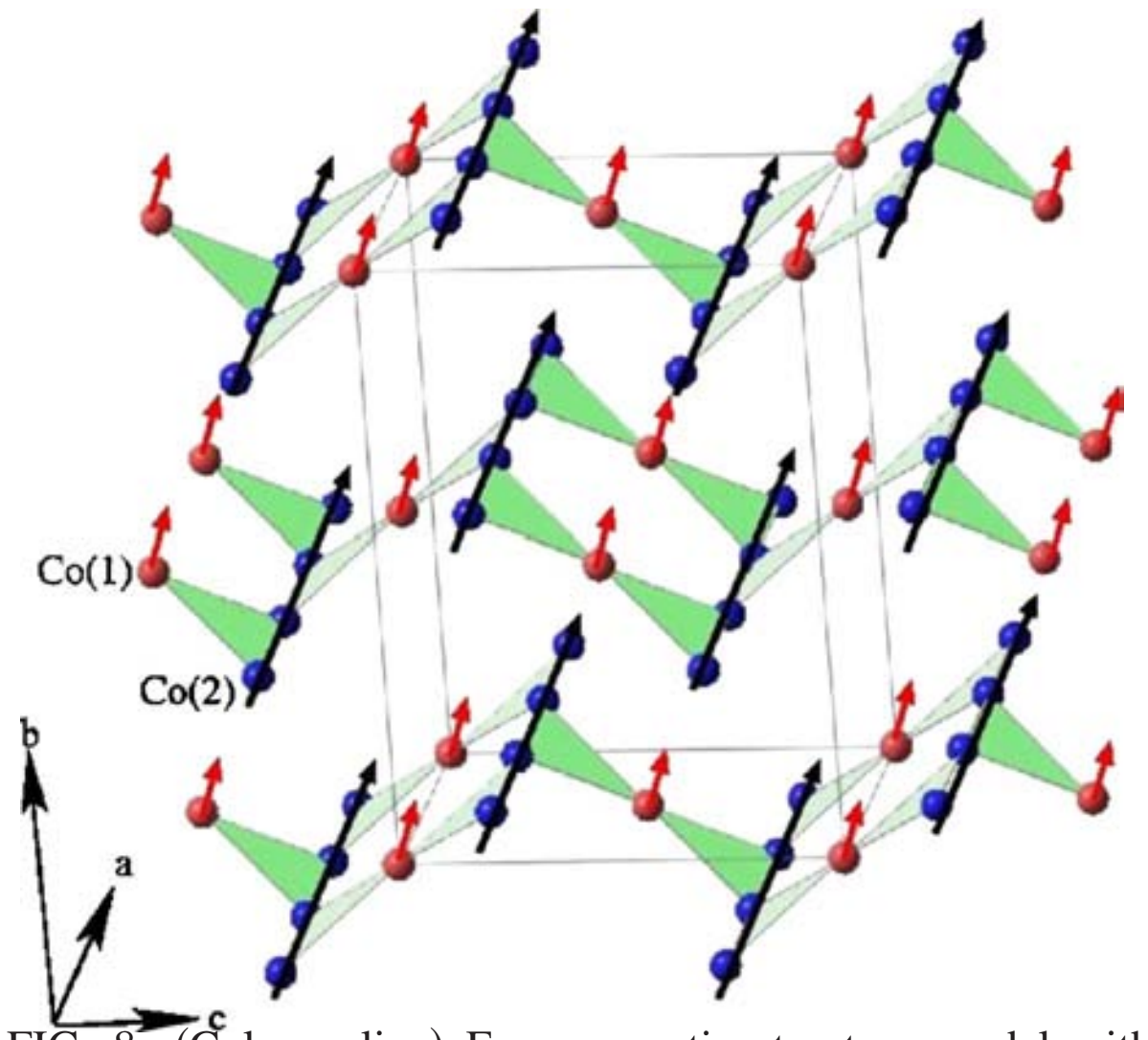

FIG. 8. (Color online) Ferromagnetic structure model with $\delta$ =0 of CVO. All the Co moments are along the crystallographic *a* axis.

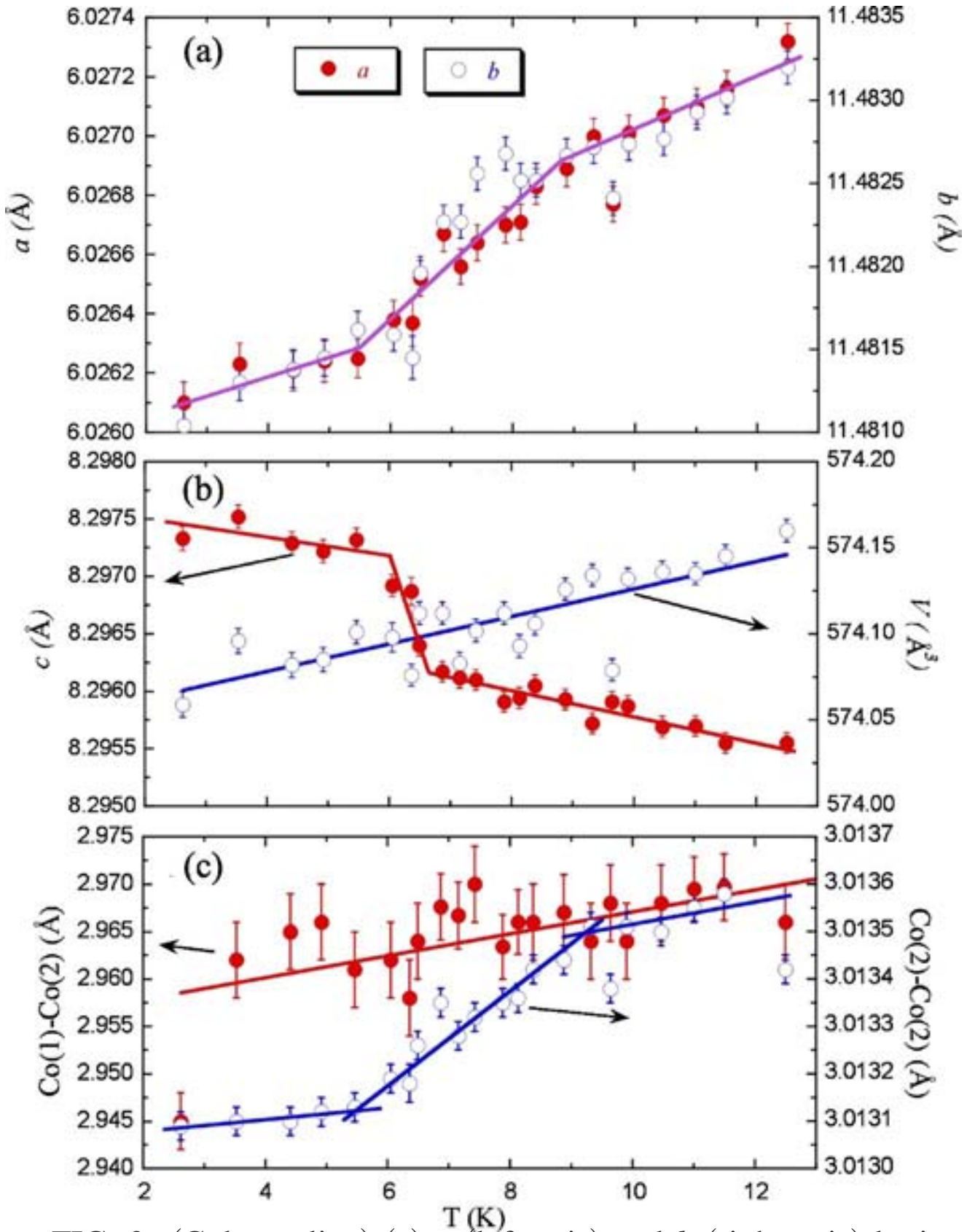

FIG. 9. (Color online) (a) *a* (left axis) and *b* (right axis) lattice parameters of CVO as a function of temperature. (b) *c* (left axis) lattice parameter and volume of the unit cell (right axis) as a function of temperature. (c) Co(1)-Co(2) distance (left axis) and Co(2)-Co(2) chain distance (right axis) as a function of temperature.

To search for possible crystal structure transitions associated with the magnetic transitions, detailed measurements of the structural parameters were made by refining the nuclear structure using the higher angular range ($2\theta > 40°$) of the neutron powder diffraction data. Figure 9 shows the lattice parameters as a function of temperatures. *a* and *b* decrease monotonically with decreasing temperature, showing a change in slope, first near the antiferromagnetic $\delta = \frac{1}{2}$ lock-in transition at 8.8 K, and then near the ferromagnetic transition at 6 K. The Co(2)-Co(2) distance in the chains along the *a* axis decreases steadily with temperature, following the behavior of the *a* axis. The *c* axis shows a negative thermal expansion with a step only near the ferromagnetic ordering temperature. The unit cell volume *V* and the distances of Co(1)-Co(2) show no anomalous changes and increase monotonically with temperature. Hence, there is no change in symmetry of the structure that can be observed with the present sensitivity when the system orders magnetically, although there is clearly some coupling through the internal coordinates of the atoms. A subtle lattice distortion is observed in single crystal polarized beam measurements as described below.

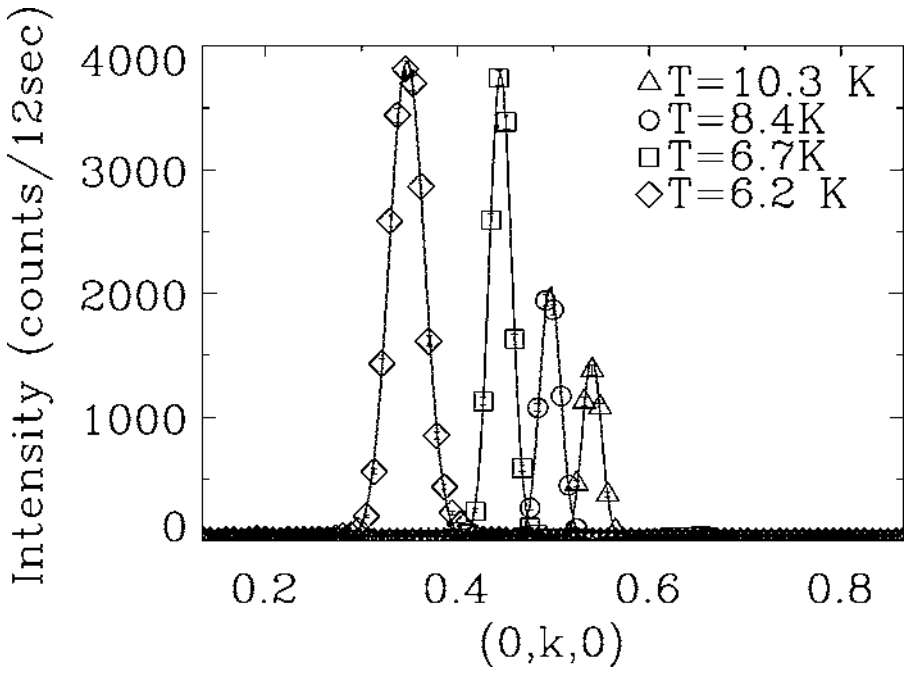

FIG. 10. $k$ scans at four different temperatures, showing that the incommensurability $\delta$ is strongly temperature dependent.

### B. Single crystal diffraction

Single crystal neutron diffraction data reveal a more complex series of incommensurate and lock-in transitions. Two reciprocal zones $(h,k,0)$ and $(0,k,l)$ were investigated at low temperatures. Magnetic peaks are only found in the $(0,k,l)$ zone, below 11.3 K. When the magnetic structure is incommensurate, the incommensurability is only along the $k$ direction.

Figure 10 shows $k$ scans along $(0,k,0)$ at four different temperatures. Intensities are observed at incommensurate position $(0,\delta,0)$ at 6.2, 6.7, 8.4, and 10.3 K with $\delta$=0.347(2), 0.452(5), 0.497(2), and 0.540(2), respectively, indicating $\delta$ is strongly temperature dependent. Such $k$ scans were measured at a series of temperatures, and a detailed temperature evolution of the incommensurate $(0,\delta,0)$ peak was determined. Figure 11 shows an intensity map as a function of $(0,k,0)$ and the temperature, upon cooling. The detailed temperature dependence of the integrated intensity, and the peak position $\delta$, are shown in Figs. 12(a) and 12(b), respectively. The data taken on cooling (circles) and warming (squares) are slightly different. The incommensurate phase begins at 11.3 K with an incommensurate peak forming at $(0,\delta,0)$ with $\delta$=0.55. With decreasing temperature, $\delta$ decreases monotonically while the integrated intensity increases gradually, before the peak locks in at the commensurate position of $\delta$=0.5 at 8.6 K. This lock-in transition marks the end of the first incommensurate phase and the beginning of a partially frustrated antiferromagnetic phase. The system remains in the $(0,\frac{1}{2},0)$ antiferromagnetic phase, with the integrated intensity being more or less constant, until 6.9 K. Below this temperature, the system moves into the second incommensurate phase. This phase is characterized by the magnetic intensity splitting into a strong reflection and a weak higher order satellite reflection (Fig. 11). With decreasing temperature, the fundamental peak $(0,\delta,0)$ shifts to lower $\delta$ and its integrated intensity changes with cooling (Fig. 12), while the weak reflection $(0,\delta',0)$ moves to higher $\delta'$, with its integrated intensity also being temperature dependent (Fig. 13). $\delta'$ can be expressed as $\delta' \approx 2-3\delta$. Alternatively, if we define the fundamental peak position as $q=\frac{1}{2}-\xi$, then the position of the higher-order satellite is given by $q' \approx \frac{1}{2}+3\xi$ [Fig. 13(c)]. The intensity of the third-order reflection is approximately proportional to that of the strong reflection. The presence of higher-order reflections indicates a squaring up of the magnetic structure, which is typical behavior for an incommensurate spin density wave system with strong uniaxial anisotropy.[29] Between 6.5 and 6.2 K, the system locks in to a commensurate $\delta=\frac{1}{3}$ phase (Fig. 12), an antiferromagnetically ordered phase that is different from the $\delta=\frac{1}{2}$ phase. This $\delta=\frac{1}{3}$ phase was not studied in detail in the powder measurements. The incommensurate intensity drops precipitously at the onset of ferromagnetic order at 6.2 K (Fig. 12). Some weak residual intensity is observed at $\delta<0.3$, but ultimately becomes unobservable below ~5.5 K.

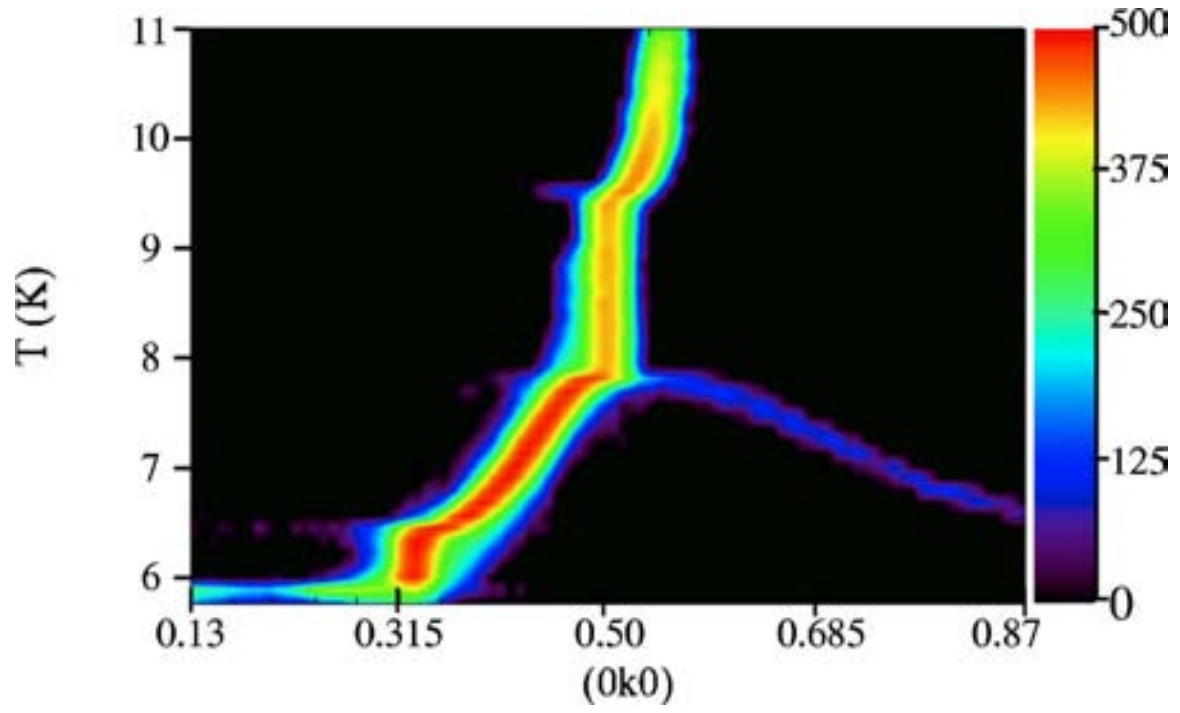

FIG. 11. (Color online) Evolution of the fundamental magnetic Bragg peak $(0,\delta,0)$ as a function of temperature, exhibiting a rich variety of incommensurate phases, interspersed with lock-in transitions to commensurate phases. The initial structure that forms is incommensurate, which locks-in to the simple frustrated antiferromagnetic structure at $\delta=\frac{1}{2}$. The fundamental peak then becomes incommensurate and moves to smaller $\delta$, while a weak higher order satellite peak moves to higher $\delta$. The structure locks in again at $\delta=\frac{1}{3}$, and finally undergoes a first-order transition to the ferromagnetic ground state ($\delta$=0).

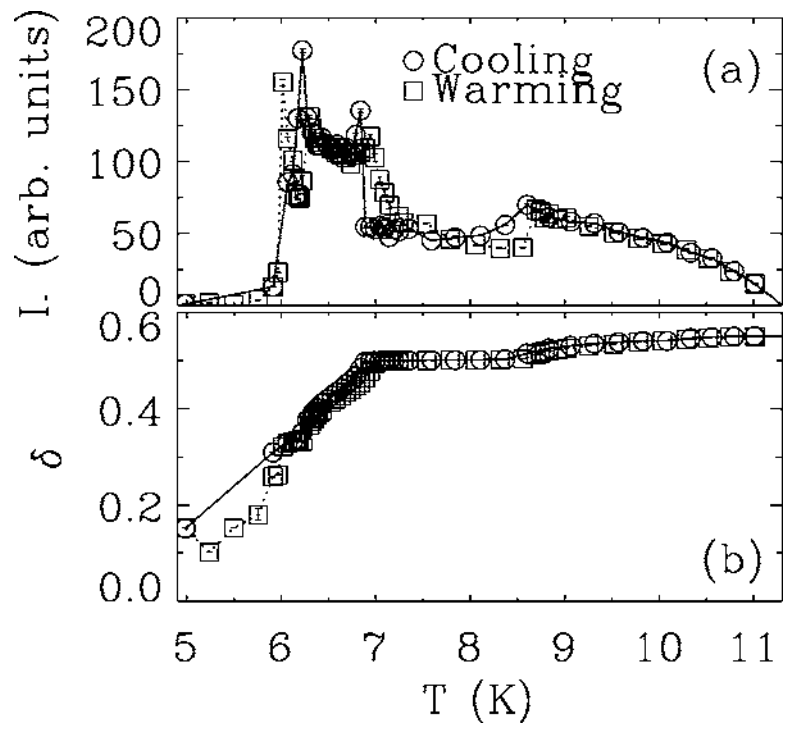

FIG. 12. (a) Integrated intensity of the fundamental incommensurate magnetic Bragg peak $(0,\delta,0)$ as a function of temperature on cooling (circles) and warming (squares). (b) Incommensurability $\delta$ as a function of temperature on cooling (circles) and warming (squares). The solid (cooling) and dotted (warming) lines in (a) and (b) are guides to the eye.

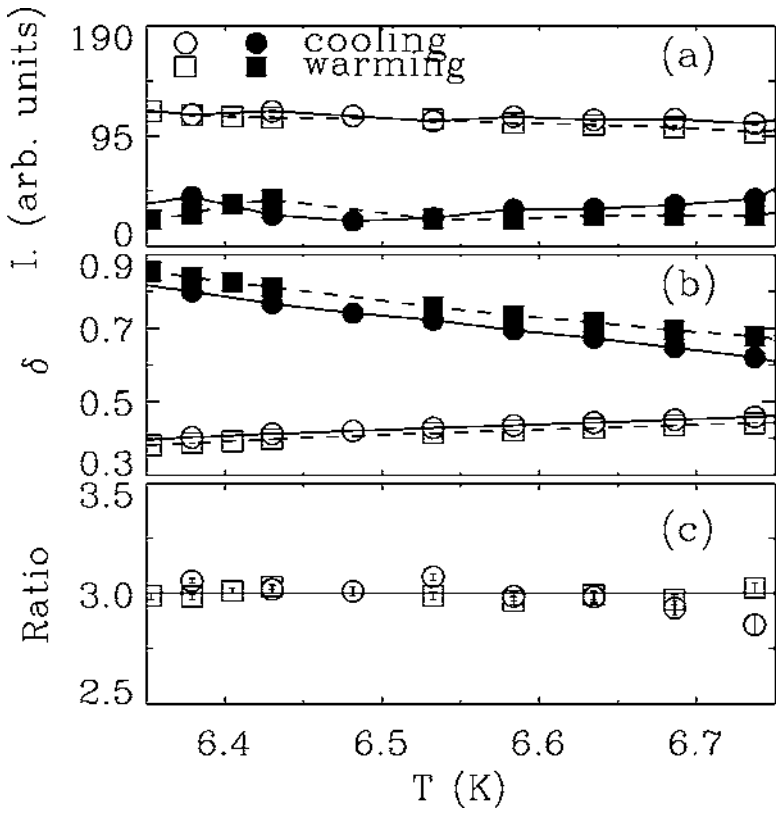

FIG. 13. (a) Integrated intensity of the (0,$\delta$,0) peak (open symbols), and 20 times the integrated intensity of the (0,$\delta'$,0) satellite peak (filled symbols) as a function of temperature on cooling (circles) and warming (squares). (b) Incommensurability $\delta$ (open symbols) and $\delta'$ (filled symbols) as a function of temperature on cooling (circles) and warming (squares). (c) Ratio $\equiv\frac{q'-0.5}{0.5-q}$ (see text) as a function of temperature. The solid (cooling) and dashed (warming) lines in (a)–(c) are guides to the eye.

Figure 14 shows the temperature dependence of the peak intensity of (0,2,1) along with that of (0,2,0) and (0,0,2) between 4 and 8.5 K on cooling (open symbols) and warming (filled symbols). The abrupt transition and the hysteresis near $T_c\approx 6.2$ K observed in all three peak intensities indicate a first-order transition from the incommensurate magnetic phase into the ferromagnetically ordered phase. In addition, the peak intensities rise sharply to a peak value at $T_c$, followed by a subsequent decay to a second value upon cooling. Such behavior, not observed in the powder sample, was also observed by polarized neutron diffraction. Figure 15 shows the ratio between spin-flip scattering and non-spin-flip scattering measured at $\mathbf{Q}=(0,2,1)$ as a function of temperature with cooling (open symbols) and warming (filled symbols), with polarization vector $\mathbf{P}\perp\mathbf{Q}$. In this configuration, the nuclear and part of the magnetic signal appear in the non-spin-flip channel. The rest of the magnetic signal appears in the spin-flip channel. A time dependence of the scattering intensities is also evident (~1 min), and its presence in the (magnetic) spin-flip channel of the polarized neutron measurements indicates that it is magnetic in origin. Such behavior is quite likely related to the formation and redistribution of magnetic domains as the ferromagnetic order develops.

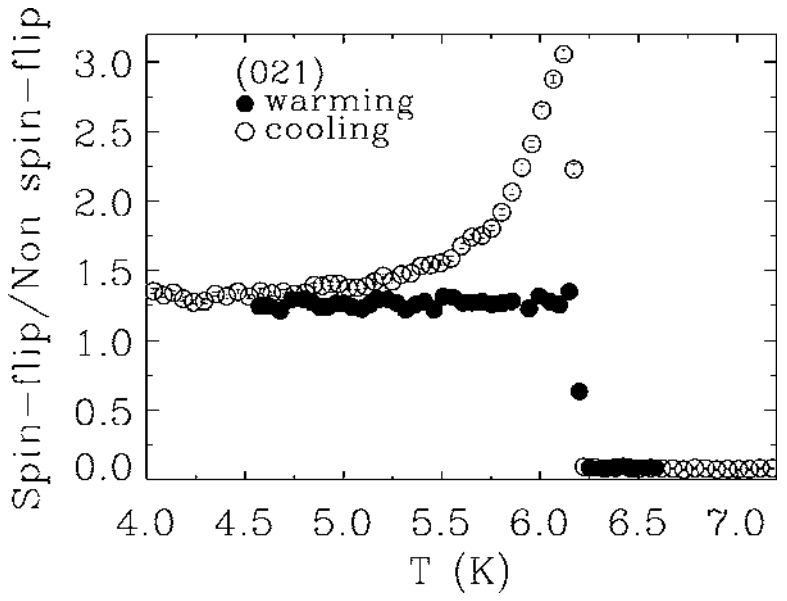

FIG. 15. Temperature dependence of the ratio between spin-flip scattering and non-spin-flip scattering at $\mathbf{Q}=(0,2,1)$, with polarization vector $\mathbf{P}\perp\mathbf{Q}$, on cooling and warming.

The antiferromagnetic phase with $\delta=\frac{1}{2}$ was investigated in detail at 8.13 K. Well-defined resolution-limited magnetic Bragg peaks are observed at $[0,(2m+1)/2,n]$, where $m$ and $n$ are integers. At 1.8 K, deep in the ferromagnetic regime, no intensities were found at $[0,(2m+1)/2,n]$.

The antiferromagnetic phase with $\delta=\frac{1}{3}$ was investigated with both unpolarized and polarized neutrons. Although the peaks of this phase occur at $\left(0,\frac{1}{3},0\right)$ and equivalent positions, we observed higher-order peaks and some of them occur at integer positions such as (0,1,0) and (0,3,0). Figure 16 shows the integrated intensity of the (0,3,0) peak as a function of temperature on cooling and warming measured with unpolarized neutrons. On warming, no intensity is observed at (0,3,0) until about 6.2 K. The observed intensity between 6.24 and 6.5 K is magnetic, associated with the $\delta=\frac{1}{3}$ antiferromagnetic phase. Interestingly, a much weaker (0,3,0) peak reappears between 7.1 and 8.7 K. The origin of this weak intensity is unknown. No intensity is observed above 8.7 K. Similar behavior is observed on cooling. Polarized neutrons were used to measure the temperature dependence of the $\mathbf{Q}=(0,1,0)$ peak on warming and on cooling with polarization vector $\mathbf{P}\parallel\mathbf{Q}$, shown in Fig. 17. In this geometry, all magnetic scattering occurs in the spin-flip channel, and all the nuclear scattering occurs in the non-spin flip channel. The non-spin-flip intensity shows a very weak peak (compared to the fundamental structural peaks) between 6.2 and 6.5 K, demonstrating that there is a crystallographic contribution to this peak. This indicates a subtle distortion of the lattice that is not apparent in the powder data. The strong spin-flip intensity observed between 6.2 and 6.5 K indicates that the dominant contribution to the (0,1,0) peak is magnetic in origin. Such behavior is observed both on cooling and on warming.

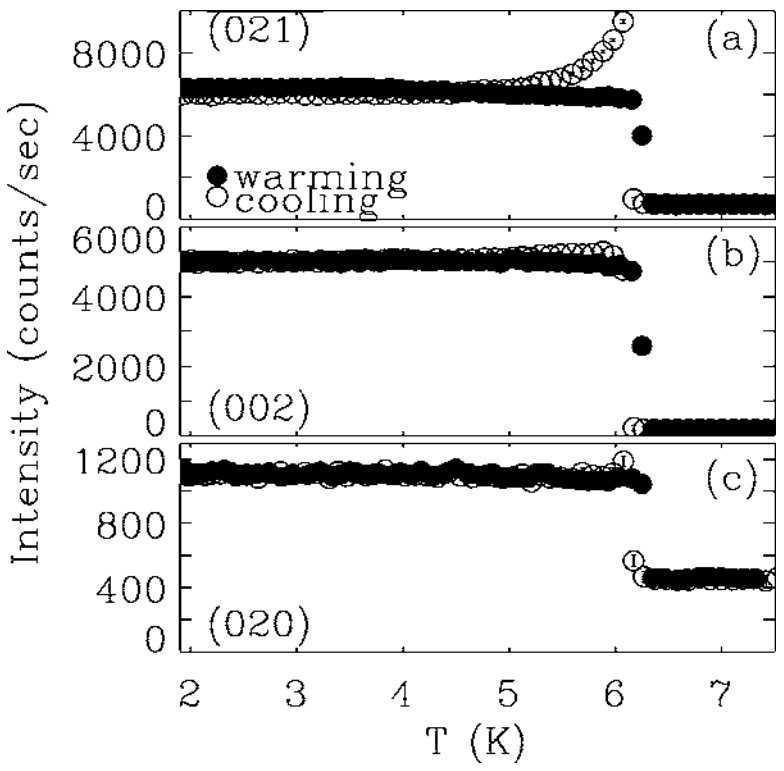

FIG. 14. Temperature dependence of the peak intensity of (a) (0,2,1), (b) (0,0,2), and (c) (0,2,0) on warming (filled circles) and on cooling (open circles).

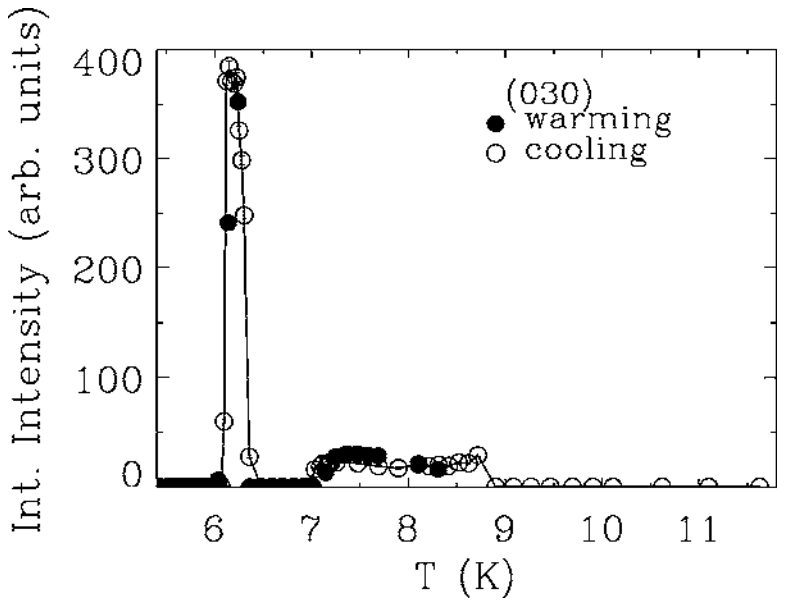

FIG. 16. The integrated intensity of the (0,3,0) peak as a function of temperature on cooling and on warming. The line is a guide to the eye.

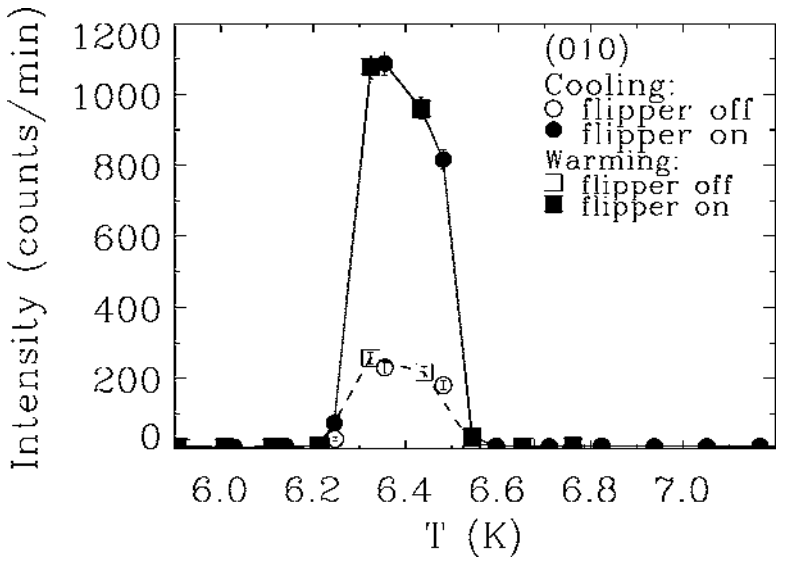

FIG. 17. Temperature dependence of the (0,1,0) peak on warming (squares) and cooling (circles) for spin-flip (filled symbols) and non-spin-flip (open symbols) channels. The solid (flipper on) and the dashed (flipper off) lines are guides to the eye.

In summary, six magnetic phases are observed in CVO: (I) the paramagnetic phase for $T>11.3$ K, (II) the first incommensurate phase with a temperature dependent propagation vector $(0,\delta,0)$ with $\delta>0.5$ for $T$ between 11.3 and 8.6 K, (III) the antiferromagnetic phase with $\delta=0.5$ for $6.9<T<8.6$ K, (IV) a second incommensurate magnetic phase with a temperature-dependent propagation vector $(0,\delta,0)$ for $6.5<T<6.9$ K, (V) a commensurate phase with $\delta=\frac{1}{3}$ for $6.2<T<6.5$ K, (VI) the ferromagnetic ground state for $T<6.2$ K.

### C. Dielectric measurements

Dielectric measurements of CVO clearly show evidence for dielectric anomalies associated with the magnetic phase transitions. Figure 18 shows the dielectric constant as a function of temperature measured on a powder sample. These data were collected at 100 kHz, but measurements at different frequencies showed qualitatively similar effects with no measurable change in transition temperature. At low temperatures, there is a dramatic drop in dielectric constant, on the order of 0.5%. By fitting straight-line segments to the $\epsilon(T)$ curve, we find that the dielectric constant begins to drop below approximately $T=6.5$ K, and then flattens out below 5.8 K. These results are consistent with either a broad dielectric anomaly associated with the $T=6.2$ K ferromagnetic transitions or distinct anomalies associated with the 6.5 and 6.2 K transitions. Higher-resolution dielectric data will be required to address this issue. This curve also shows a small feature at $\sim$11 K (see dashed lines in Fig. 18), which is consistent with observations of a phase transition at this temperature in both specific heat and neutron measurements. This temperature dependence of the dielectric constant strongly suggests that there is significant spin-charge coupling present in CVO. Preliminary investigations on single crystal samples establish that the anomaly is only observed for $\vec{E}\|$ (001), and there is no significant dielectric anomaly measured along the (100) and (010) direction. A moderate magnetic field of 2 T suppresses this anomaly completely.

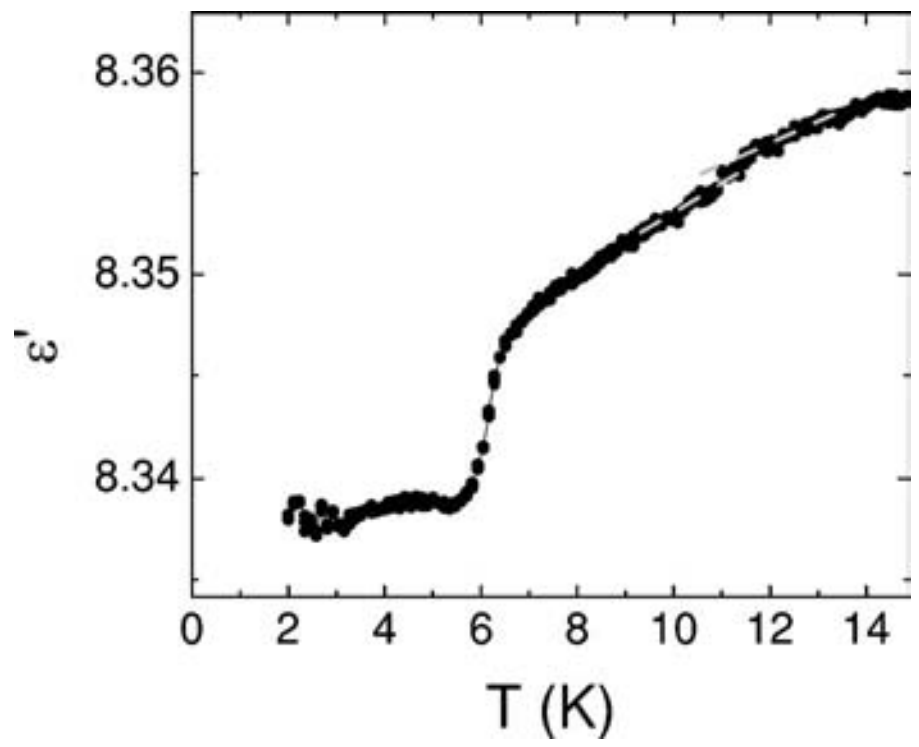

FIG. 18. Dielectric constant as a function of temperature of a powder sample of CVO measured at 100 K Hz. The dashed lines are guides to the eye.

## IV. THEORY

Details of the theory will be presented elsewhere.[30] Here we only give a brief outline of the theory, which gives a qualitative explanation of the incommensurate and commensurate magnetic structures. The theory contains two main parts, based on a group theory analysis and on a minimal magnetic interaction model.

Since the atomic structure of CVO is practically the same as that of NVO, both systems can be analyzed with a similar basic group theory analysis, as presented in Ref. 31. However, since CVO exhibits commensurate and incommensurate order with the wave vector along the $b$ axis (whereas NVO has wave vectors along the $a$ axis), CVO requires an analysis of the irreducible representations (irreps) that correspond to a wave vector along $\mathbf{b}$. The relevant symmetry operations include $\sigma_x$, that is a reflection through a plane that is perpendicular to the $a$ axis and passes through a cross-tie Co(1) site, and $2_y$, which is a twofold rotation about the $b$ axis passing through a spine Co(2) site (see Fig. 2). The resulting irreps are labeled as $\Gamma_{s_1,s_2}$, according to the eigenvalues $\{s_1,s_2\}$ of $\{\sigma_x,2_y\}$. Out of the four irreps, only $\Gamma_{+-}$ seems to be consistent with the observed data.

A detailed analysis of the constraints imposed by the irrep $\Gamma_{+-}$ then allows us to parametrize the spins in the unit cell in the following form:

$$\mathbf{m}_{\ell,1}=[a\cos(\mathbf{q}\cdot\mathbf{R}+\phi+\alpha),\ 0,\ b\cos(\mathbf{q}\cdot\mathbf{R}+\phi+\beta)],$$

$$\mathbf{m}_{\ell,2}=[a\cos(\mathbf{q}\cdot\mathbf{R}+\phi+\alpha),\ 0,\ -b\cos(\mathbf{q}\cdot\mathbf{R}+\phi+\beta)],$$

$$\mathbf{m}_{u,1}=[a\cos(\mathbf{q}\cdot\mathbf{R}+\phi-\alpha),\ 0,\ -b\cos(\mathbf{q}\cdot\mathbf{R}+\phi-\beta)],$$

$$\mathbf{m}_{u,2}=[a\cos(\mathbf{q}\cdot\mathbf{R}+\phi-\alpha),\ 0,\ b\cos(\mathbf{q}\cdot\mathbf{R}+\phi-\beta)],$$

$$\mathbf{m}_c = [c\cos(\mathbf{q}\cdot\mathbf{R}+\phi),\ 0,\ 0]. \tag{1}$$

Here, each unit cell within a buckled plane contains two neighboring Co(2) spine spins on the "lower" spine chain, denoted by $\{\ell,1\}$ and $\{\ell,2\}$, two similar Co(2) spins on the "upper" spine chain, denoted by $\{u,1\}$ and $\{u,2\}$, and two cross-tie Co(1) spins, which we denote simply by the subscript $c$. The vector $\mathbf{R}$ denotes the unit cell. For a general incommensurate vector $\mathbf{q}=(0,\pi\delta,0)$, we thus need real parameters, namely, $a$, $b$, $\phi$, $\alpha$, and $\beta$. As we discuss below, locked-in commensurate wave vectors imply additional relations between the phases $\alpha$, $\beta$, and $\phi$. The magnetic structure is illustrated in Fig. 7 for the special case $\delta=1/2$. Note that by suitably choosing the phases $\phi$ and $\alpha$ in Eq. (1), we can reproduce the different moments shown in Fig. 7 for the two different Co(2) sites (see Sec. III A). Within the accuracy of the experimental data presented above, there appear no spin components along the $c$ axis. We hope that future data will allow a full confirmation of the predictions in Eq. (1). The $\mathbf{c}$ components of the spins could arise from Dzyaloshinskii-Moriya and pseudodipolar interactions,[31] to be discussed elsewhere.[30] However, apparently, the spin anisotropy in CVO is large, forcing the ordered moments to be mostly along the $a$ axis. For simplicity, we therefore assume from now on that the spins order only along the $a$ axis, namely, that $b=0$ in Eq. (1). Under this assumption, all the spins along each spine chain are equal to each other, probably due to a relatively strong Ising-like ferromagnetic nearest-neighbor interaction [denoted by $J_2$ in Fig. 2(a)]. For the $n$th buckled layer, we thus denote the $\mathbf{a}$ components of the (lower and upper) alternate spine chains by $S_\ell(n)$ and $S_u(n)$. To obtain an incommensurate order, the effective spin Hamiltonian needs to contain competing interactions among spins along the $b$ axis. A minimal such Hamiltonian, only for the spine spins, has the form[30]

$$\mathcal{H} = -\sum_n \{S_u(n)[2J_{u\ell}S_\ell(n+1) + 2J_{\ell u}S_\ell(n-1) + 2J_3S_\ell(n) + J_4S_u(n+1)] + J_4S_\ell(n)S_\ell(n+1)\}, \tag{2}$$

where $J_{u\ell}$ ($J_{\ell u}$) denotes the exchange interaction between an "upper" ("lower") spine spin and the "lower" ("upper") next-nearest spine spin in the layer above it. Similarly, $J_3$ denotes the coupling between nearest-neighbor spine chains in a specific buckled plane, while $J_4$ denotes the coupling between neighboring spine chains in neighboring buckled planes (i.e., chains above and below each other in the $b$ direction). All these interactions are labeled in Fig. 2.

A Landau expansion in terms of the Fourier transforms $S_u(q)$ and $S_\ell(q)$, yields a wide range of parameters where one finds an incommensurate optimal value $q=\pi\delta$, for which the Ising spin order parameters obey Eq. (1) with $b=0$. In particular, this value of $q$ turns out to depend strongly on the intraplane coupling $J_3$ (which effectively varies with temperature, see below): For a large positive $J_3$ the model yields a ferromagnetic ordering of all the spine chains. However, as $J_3$ decreases, one enters a regime where $\delta$ increases monotonically from 0 toward 1, which corresponds to a full antiferromagnetic order within all planes. Apparently, the data indicate a large positive effective $J_3$ at low temperature $T$, which decreases with increasing $T$.

To obtain the temperature dependence of $J_3$, we next add the spine–cross-tie coupling, $J_1$ in Fig. 2(a). Writing the corresponding Hamiltonian as

$$\mathcal{H}_{cs} = -2J_1S_c(n)[S_\ell(n) + S_u(n)], \tag{3}$$

and assuming a relatively small susceptibility of the cross-tie spins, we can integrate $S_c(n)$ out of the problem, generating an effective spine-spine coupling,

$$\Delta\mathcal{H} = -2\chi_cJ_1^2\sum_n [S_\ell(n) + S_u(n)]^2. \tag{4}$$

This coupling implies a temperature-dependent shift in the effective intraplane coupling $J_3$,

$$J_3 \to J_3 + 2\chi_cJ_1^2. \tag{5}$$

Assuming that $\chi_c$ increases with decreasing temperature $T$, this variation of $J_3$ with $T$ can explain the observed decrease of $q$ with decreasing $T$.[30] We remark that a similar temperature dependence of the interactions is responsible for the spin-flop transitions in $Nd_2CuO_4$,[32,33] and may be responsible for the spin-flop transition in $HoMnO_3$ as well.[9]

To explain the higher harmonics and the locked-in commensurate structures, we need higher-order terms in the Landau expansion. Specifically, terms like $S_u(q)^3S_u(2\pi-3q)$ will generate spins with wave vectors $\delta'=2-3\delta$. Also, the terms $[S_u(\frac{\pi}{2})^4+S_\ell(\frac{\pi}{2})^4+cc]$ favor commensurate order with $\delta=\frac{1}{2}$, over a range of parameters. For this locked-in solution, one also obtains the restriction $\phi=2\alpha$ (or $\phi=2\alpha+\frac{\pi}{4}$, depending on specifics.[30]) For the data presented in Fig. 7, this implies that $\phi=0$ and $\alpha\approx-0.34\pi$.

Similarly, the term $[S_u(\frac{\pi}{3})^6+S_\ell(\frac{\pi}{3})^6+cc]$ generates a locked-in structure with $\delta=\frac{1}{3}$, with similar restrictions on the phases $\phi$ and $\alpha$. We hope to test these predictions in future work.

The transitions into the locked-in phases are all predicted to be first order. Apparently, at low temperature there appears a direct first-order transition from the phase with $\delta=\frac{1}{3}$ to the ground state ferromagnetic phase. At these temperatures, our Landau expansion is only qualitative, due to many higher harmonics. Lowering $T$ from the paramagnetic phase, one first encounters a continuous phase transition into an incommensurate phase with a wave vector that corresponds to some intermediate value of $J_3(T)$.

## V. DISCUSSION

The crystal symmetry of CVO and NVO is identical, and the structure parameters are very similar. Hence, one might have expected that CVO and NVO would have quite similar magnetic and multiferroic properties. This is not the case because these systems are nearly fully frustrated magnetically. This frustration is relieved by perturbations, which, in most contexts, would be irrelevant but, here, determine the general aspects of the magnetic structure. In both NVO and CVO, the spins mainly order along the $a$ axis. Apparently, this uniaxial anisotropy is somewhat stronger in CVO. A major difference concerns the ordering along each spine. In NVO, a competition between the nearest- and next-nearest-neighbor interactions yields incommensurate ordering along the spine, ending with an antiferromagnetic order at low temperatures. In contrast, for CVO the spins along each spine order ferromagnetically at all temperatures. In NVO, incommensurability arose because the Ni(2)-O-Ni(2) bond angle for the nearest-neighbor superexchange interaction alternated between 90.4° and 95.0°, in a region where this interaction changes quickly from being ferromagnetic (at 90°) to antiferromagnetic (as the angle moves away above 90°). As a result, the small antiferromagnetic nearest-neighbor interaction was comparable to the next-nearest-neighbor interaction.

Once we understand the major difference in the ordering of a single spine, it is easy to see why other interactions, such as those between intra- and interplanar neighboring spines, play a more important role in CVO. In NVO, neighboring planes order antiferromagnetically, while in CVO one encounters delicate competitions between various neighbors, as described by our minimal theoretical model. It would be nice to have an explicit derivation of all these interactions (as well as of the spin anisotropies), based on some microscopic model.

As seen in Sec. IV, a minimal Ising model with four competing spine-spine exchange interactions, including a temperature-dependent interaction $J_3$ between neighboring spines, which is mediated by the cross-tie spins, is sufficient to capture all the qualitative features observed experimentally. At low temperature, this effective interaction $J_3$ is ferromagnetic, locking all the spins into a ferromagnetic structure. As $T$ increases, $J_3$ decreases, ending up with an incommensurate structure, with a wave vector along the $b$ axis. The wave vector $\delta$ increases gradually with decreasing $J_3$, locking-in at the commensurate values $\frac{1}{3}$ and $\frac{1}{2}$ due to higher-order terms in the Landau expansion and disappearing at $\delta \approx 0.55$ above $T = 11.3$ K. Our theory also allows some spine spin components along the $c$ direction and predicts several relations among the parameters, which describe the spin structures. Some of these predictions still await future experimental verification.

CVO is one of the rare systems that exhibits a coupling of the ferromagnetic magnetic order with the dielectric properties, which may render it a paradigm for systems of interest to both the multiferroic and spintronics communities. Our results also highlight the importance of characterizing these multifunctional oxides using a range of different experimental techniques. Although many different physical properties exhibit anomalies at phase transitions, certain quantities are much more strongly affected than others. By investigating the response using a variety of techniques, it is possible to develop a deeper understanding of the materials couplings underlying the transitions.

## ACKNOWLEDGMENTS

We thank W. Ratcliff and O. P. Vajk for useful discussions. We acknowledge support from the US-Israel Binational Science Foundation (BSF). Work at the University of Maryland was supported in part by NSF-MRSEC, Grant No. DMR 05-20471. The research at Princeton was supported by the NSF Solid Sate Chemistry Program.